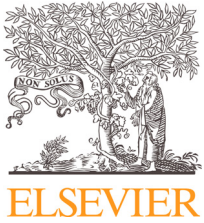
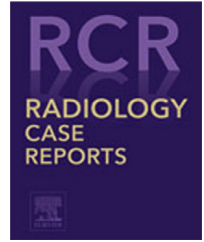

Case Report

# Quantitative analysis of diaphragm motion during fluoroscopic sniff test to assist in diagnosis of hemidiaphragm paralysis☆

*Jacky Chow, MD, MBA, PhD\*, Muhammed Hatem, MD*

*Department of Radiology, Cumming School of Medicine, University of Calgary, 3330 Hospital Dr NW, Calgary, Alberta, T2N 4N1, Canada*

ABSTRACT

The current imaging gold standard for detecting paradoxical diaphragm motion and diagnosing hemidiaphragm paralysis is to perform the fluoroscopic sniff test. The images are visually examined by an experienced radiologist, and if one hemidiaphragm ascends while the other descends, then it is described as paradoxical motion, which is highly suggestive of hemidiaphragm paralysis. However, diagnosis can be challenging because diaphragm motion during sniffing is fast, paradoxical motion can be subtle, and the analysis is based on a 2-dimensional projection of a 3-dimensional surface. This paper presents a case of chronic left hemidiaphragm elevation that was initially reported as mild paradoxical motion on fluoroscopy. After measuring the elevations of the diaphragms and modeling their temporal correlation using Gaussian process regression, the systematic trend of the hemidiaphragmatic motion along with its stochastic properties was determined. When analyzing the trajectories of the hemidiaphragms, no statistically significant paradoxical motion was detected. This could potentially change the prognosis if the patient was to consider diaphragm plication as treatment. The presented method provides a more objective analysis of hemidiaphragm motions and can potentially improve diagnostic accuracy.

## Introduction

Chronic unilateral diaphragm paralysis is an uncommon and underdiagnosed cause of dyspnea with an unknown incidence [1,2]. The fluoroscopic sniff test is often considered the imaging gold standard for diagnosing unilateral diaphragm paralysis [2]. Recent research has identified paradoxical motion as a favorable prognosis factor in patients pursuing diaphragm plication [3]. Conventionally, paradoxical motion is detected on fluoroscopy based on a radiologist's visual analysis and experience, which can be challenging at health care facilities where the test is not frequently performed. We present a case of possible mild paradoxical motion on fluoroscopic sniff test, where

☆ Competing Interest: The authors declare no conflict of interest or financial disclosures.
\* Corresponding author.
E-mail address: jckchow@ucalgary.ca (J. Chow).
https://doi.org/10.1016/j.radcr.2022.02.083

we measured the elevations of the hemidiaphragms and estimated their trajectories using Gaussian process regression. Quantitative analysis of the diaphragmatic trajectory provides additional information to assist in the imaging diagnosis.

## Case report

The patient is a 57-year-old female originally from Pakistan. Her relevant medical history includes type-2 diabetes, gastroesophageal reflux, obesity (BMI 32), and lifelong asthma (diagnosed in childhood) that is well controlled on a tiotropium bromide inhaler (Spriva 18 mcg daily), fluticasone furoate/vilanterol trifenatate inhaler (Breo Ellipta 200-25 mcg daily), montelukast sodium (Singulair 10 mg daily), and salbutamol sulfate inhaler (Ventolin 100 mcg) as needed. On average, she has been requiring 2-3 courses of prednisone every year.

She was never informed of any diaphragm abnormalities growing up. After a cesarean section under spinal anesthesia in 2000 for a singleton, she developed post-operative ileus requiring nasogastric decompression. Since the cesarean section, she has experienced worsening dyspnea. In the following year (ie, 2001), at the age of 37, she was diagnosed with left hemidiaphragm elevation from unknown etiology on chest radiograph for the first time. She denies any trauma or other surgeries, as well as any chest pain, orthopnea, or dysphagia. Her dyspnea has been stable for the past 20 years, and the degree of hemidiaphragm elevation is unchanged since.

Her most recent pulmonary function test in 2021 demonstrated a restrictive pattern with FEV1/FVC ratio of 79%, FEV1 1.30 L (58% predicted), FVC 1.65 L (59% predicted), and DLCO 13.91 mL/min/mmHg (73% predicted). Chest CT demonstrated marked elevation of the left hemidiaphragm with no obvious masses and/or underlying pulmonary intraparenchymal disease. Upper gastrointestinal and small bowel follow-through fluoroscopy showed no significant impairment in her foregut. Diaphragm fluoroscopy showed reduced movement of the left hemidiaphragm when compared to the right with the possible subtle paradoxical movement of the left hemidiaphragm.

She was offered diaphragm plication in 2002, but she declined. At her most recent thoracic consult in 2021, the surgeon was not convinced that the findings of the left hemidiaphragm were related to the cesarean section in the year 2000. Her chronic left hemidiaphragm elevation could be post-viral or idiopathic; the exact cause is still unknown.

## Discussion

The diaphragm has two main functions: (1) anatomically, it separates the thoracic cavity from the abdominal cavity, and (2) functionally, it is the main respiratory muscle that contracts and descends during inspiration to increase negative intrathoracic pressure, which draws air into the lungs. Although the diaphragm is less commonly considered part of the gastrointestinal and urinary systems, it assists in defecation and urination by increasing the intra-abdominal pressure [4].

Diaphragm dysfunction can be either unilateral (one diaphragm is affected) or bilateral (both diaphragms are affected). There are three main types of dysfunctions in the diaphragm: paralysis, weakness, and eventration [4,5]. Paralysis is described as the absence of downward diaphragm motion during normal breathing with paradoxical motion (ie, upward diaphragm motion) when sniffing. Weakness is defined as reduced/delayed downward diaphragm motion during normal breathing, with or without paradoxical motion. Eventration is often congenital and is described as focal thinning/weakening of the diaphragm. It can typically be separated from the other two classifications using cross-sectional anatomical imaging, for example, computed tomography (CT).

The motor function of the diaphragm is innervated by the phrenic nerve, which originates from the C3, C4, and C5 spinal nerves in the cervical spine. Trauma or disease process anywhere along the respiratory chain from the brainstem to the diaphragm can result in functional impairment; possible pathologies include central nervous system lesions such as brain stem or cervical cord tumor and amyotrophic lateral sclerosis, peripheral nerve dysfunction such as Guillain-Barre syndrome or chronic inflammatory demyelinating polyneuropathy, myopathies such as systemic lupus erythematous, and phrenic nerve dysfunction such as post-viral or idiopathic phrenic neuropathy [5,6]. Dysfunction of the neurogenic control system is a common cause of diaphragm paralysis. Abnormalities in the diaphragm muscles causing paralysis are rare [5]. If the paralysis is severe, patients can develop significant dyspnea that reduces their daily function and quality of life [3].

Various modalities have been utilized clinically to assess for diaphragm paralysis. Bilateral diaphragm paralysis is a challenging diagnosis as it often presents similar to severe diaphragm weakness and fatigue [5]. Symptoms are often more severe than unilateral diaphragm paralysis. To make a reliable diagnosis, invasive transdiaphragmatic pressure measurement with gastric and esophageal balloons is often necessary. If for instance, the cause is secondary to phrenic nerve dysfunction, EMG and nerve conduction studies can be considered, but these diagnostic tools are rather invasive [6]. Fortunately, bilateral diaphragm paralysis is uncommon; most cases of diaphragm paralysis are unilateral [1,4].

This paper focuses on the diagnosis of unilateral diaphragm dysfunction using X-ray fluoroscopy by proposing/discussing the use of quantitative measures derived from image processing. Often, unilateral diaphragm dysfunction is first suspected when one diaphragm is incidentally identified as abnormally elevated on chest radiograph (Fig. 1). However, static analysis of the height of one hemidiaphragm relative to the contralateral side does not appear to correlate with surgical outcomes the way dynamic imaging does [3,7]. In addition, the sensitivity of chest radiograph for detecting unilateral diaphragm paralysis is reported as only 66.6% [8]. Further diagnostic workups often include a pulmonary function test (PFT) and dynamic imaging, such as ultrasound or fluoroscopy. In patients with diaphragm paralysis, PFT often demonstrates a restrictive pattern where the forced vital capacity (FVC) and forced expiratory volume in 1 second (FEV1) are reduced. However, this is a global assessment of ventilation, and does not distinguish diaphragm dysfunction from other causes of re-strictive patterns (eg, pleural effusion), left diaphragm dysfunction from right, or diaphragm paralysis from weakness. Therefore, radiologic study is often ordered in conjunction to assess for anatomical and functional abnormalities.

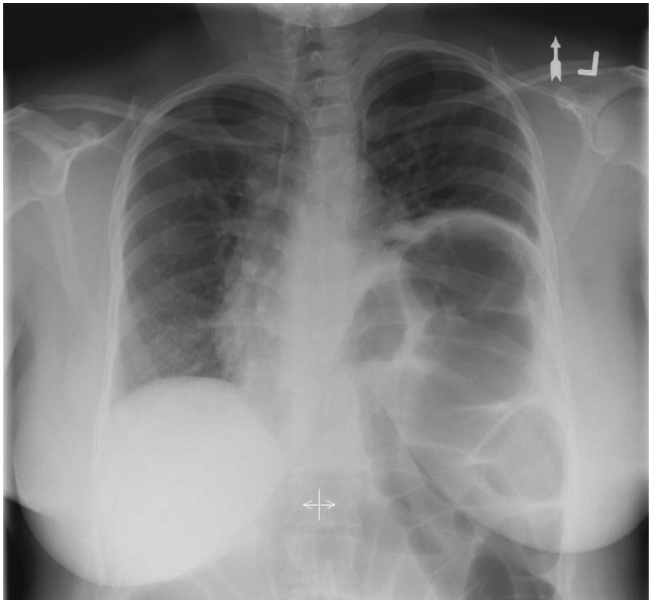

Fig. 1 – Posterior-anterior projection upright chest radiograph of the patient with chronic left hemidiaphragm elevation from unknown etiology.

Traditionally, diaphragm fluoroscopy (a.k.a. fluoroscopic sniff test) is considered the imaging gold-standard for diagnosing unilateral diaphragm paralysis [8]. It has a diagnostic sensitivity of 90% [6], but poor specificity [9]. During normal inspiration both diaphragms contract and descend; during the expiratory phase, the diaphragms relax and ascend. This sinusoidal pattern repeats every respiratory cycle and is typically regulated by the autonomic nervous system. In healthy individuals, it is not uncommon to have mild asymmetry in hemidiaphragm motion and mild temporal offsets [10]. To be considered paradoxical motion, a 180-degree phase difference would be expected between the signal of the pathological side and the contralateral side. The absolute magnitude of diaphragmatic motion during normal breathing varies significantly between individuals based on their sex, age, body habitus, etc., and ranges from 1 to 2.5 cm [8,11]. During the sniffing portion of the study, any paradoxical motion tends to be accentuated (ie, increased in magnitude).

In current clinical practice, separating diaphragm paralysis from weakness is not critical and may not change clinical management. If a patient has asymmetric diaphragm paralysis that is acquired, chronic, and symptomatic, a thoracic surgeon could consider diaphragm plication as a form of surgical treatment. However, recent research has demonstrated that paradoxical motion is a good predictor for positive surgical outcomes, which makes sense because negative work is exerted at every breath [3]. Therefore, identifying the presence of paradoxical motion has important clinical value. The challenges are that the sniffing motion is a high-frequency signal, and the classification of paradoxical motion vs no paradoxical motion is determined qualitatively by the radiologist interpreting the imaging series.

To address the first issue, fluoroscopy data should be captured at high frequencies (eg, 30 frames per second or higher) to reduce motion blur. Unlike ultrasound-based assessment of diaphragm paralysis [8,12], fluoroscopy allows both diaphragms (ie, the presumed normal side and the dysfunctional side in unilateral paralysis) to be acquired simultaneously for comparison, and the full contour of the diaphragm is better assessed than in profile mode by ultrasound.

For quantitative diagnosis of paradoxical motion, the heights of the diaphragmatic domes during quiet inspiration/expiration and sniffing can be measured and plotted over time (Fig. 2). In this paper, all image processing was performed using MATLAB. We employed the Canny edge detector to delineate the diaphragm contour in every frame. An engineer (with over 10 years of image processing experience) then manually selected a stable/reproducible edge pixel near the dome in each frame to represent the height of the diaphragm. The forward and backward spatial/temporal correlation, as well as individual measurement noise, was then taken into consideration by modeling the data using Gaussian process regression with a squared exponential kernel. The smoothed motion of each hemidiaphragm along with its corresponding standard deviation was then estimated and plotted for graphical analysis (Fig. 2). From the time series, the amplitude and frequency/period of the diaphragms can be estimated by computing the best-fit sinusoidal function that minimizes the least-squares error (Table 1). When comparing the hemidiaphragms side-to-side for the patient described in this case, we used least-squares estimation to determine the phase offset and amplitude scale factor between the two modeled trajectories by matching the automatically detected peaks and troughs.

During quiet breathing, the hemidiaphragms were moving at a statistically different amplitude based on the t-test at a 95% confidence interval (P-value < .001). The left hemidiaphragm was moving with an amplitude that was less than half of the right (ie, 41.0% +/- 0.2% the amplitude of the right). On the other hand, there was no statistically significant difference in the signal frequency of the hemidiaphragms at the 95% confidence interval (P-value = .94); the phase offset of the left hemidiaphragm compared to the right was 18.2 deg +/- 11.2 deg (this represents a delay in movement by 0.10 seconds +/- 0.06 seconds).

Similarly, during rapid sniffing, there was also no evidence of paradoxical motion (Fig. 2B). The amplitude of the left hemidiaphragm was 36.6% +/- 12.4% that of the right, which was considered significantly different at the 95% confidence interval (P-value = .03). Although the frequencies of the hemidiaphragms were also significantly different at the 95% confidence interval (P-value = .02), the estimated phase offset of the left hemidiaphragm compared to the right was 59.5 deg +/- 25.9 deg (which corresponds to a time delay of 0.05 seconds +/- 0.02 seconds). This is far from the 180 degrees that would be classified as paradoxical motion (P-value < .001).

In the presented case, the radiologist reported the fluoroscopic sniff test as positive for diaphragm paralysis be-cause the mild paradoxical motion was visually observed. In retrospect, as seen in Figure 2, when the trajectories of the diaphragmatic domes were measured and plotted side-by-side, there is no paradoxical motion (ie, no significant hemidiaphragm motion with a phase offset of 180 degrees). The motion of the left hemidiaphragm had reduced amplitude and was slightly delayed when compared to the right (but within normal limits), which is more suggestive of hemidiaphragmatic weakness rather than paralysis. If this information was available at the time of reporting, it could have potentially changed the radiologist's interpretation of the data, which could alter the report and have a stronger correlation with the prognosis and expected benefits if the patient was to pursue diaphragm plication as a form of treatment.

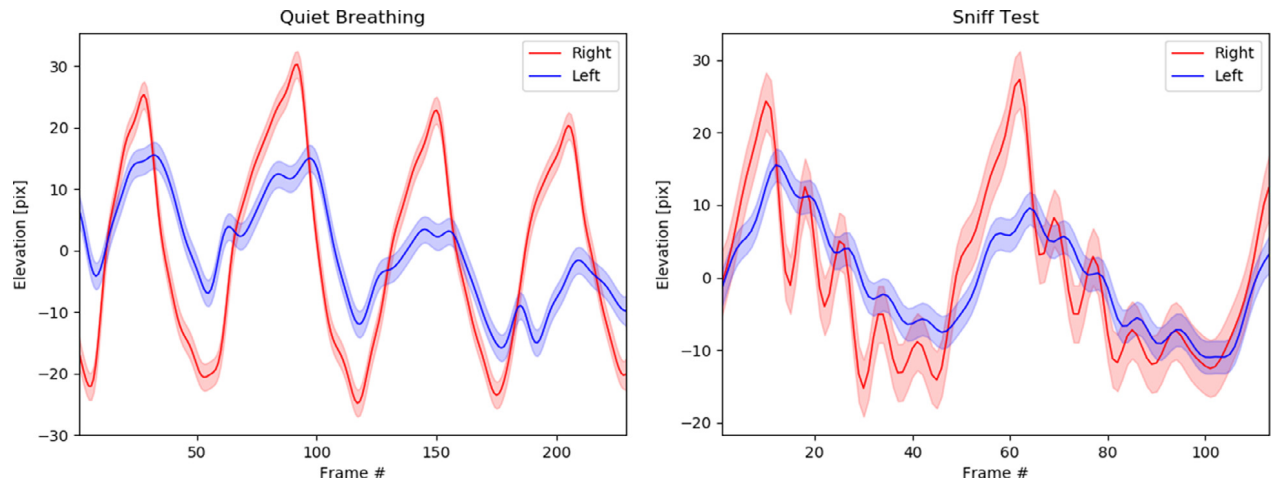

Fig. 2 – (A) Diaphragm elevation motion during quiet inspiration and expiration of the patient with chronic left hemidiaphragm elevation. The left diaphragm is weaker and has reduced amplitude in both inspiratory and expiratory phase (ie, amplitude scale difference of 41.0% +/- 0.2%). There is also a mild phase offset between the two hemidiaphragm, with the left lagging behind the right by 0.10 seconds +/- 0.06 seconds. (B) Diaphragm elevation during sniff test demonstrating that the two hemidiaphragm are mildly out of synchronization (but within normal limits) with a phase delay of 0.05 seconds +/- 0.02 seconds and decreased amplitude of the left hemidiaphragm trajectory when compared to the right (measuring 36.6% +/- 12.4%), but no evidence of paradoxical motion to suggest paralysis. Note: the color bounds represent the 95% confidence interval.

| Table 1 – Wave characteristics of the diaphragms. | | | | |
|---|---|---|---|---|
| | Quiet breathing | | Sniffing | |
| | Left | Right | Left | Right |
| Amplitude [pix] | 9.3 +/- 5.9 | 23.7 +/- 3.2 | 2.6 +/- 1.2 | 7.7 +/- 4.2 |
| Period [s] | 1.92 +/- 0.28 | 1.93 +/- 0.20 | 0.33 +/- 0.04 | 0.27 +/- 0.02 |

## Consent

Written informed consent has been obtained from the patient(s) for this publication.

REFERENCES

[1] Qureshi A. Diaphragm paralysis. Semin Respir Crit Care Med 2009;30(3):315–20. doi:10.1055/s-0029-1222445.
[2] Ricoy J, Rodríguez-Núñez N, Álvarez-Dobaño J, Toubes M, Riveiro V, Valdés L. Diaphragmatic dysfunction. Pulmonary 2019;25(4):223–35. doi:10.1016/j.pulmoe.2018.10.008.
[3] Patel D, Berry M, Bhandari P, Backhus L, Raees S, Trope W, et al. Paradoxical motion on sniff test predicts greater improvement following diaphragm plication. Ann Thorac Surg 2021;111(6):1820–6. doi:10.1016/j.athoracsur.2020.07.049.
[4] Nason L, Walker C, McNeeley M, Burivong W, Fligner C, Godwin J. Imaging of the diaphragm: anatomy and function. RadioGraphics 2012;32(2):E51–70. doi:10.1148/rg.322115127.
[5] Tarver R, Conces D, Cory D, Vix V. Imaging the diaphragm and its disorders. J Thorac Imaging 1989;4(1):1–18.
[6] Billings M, Aitken M, Benditt J. Bilateral diaphragm paralysis: a challenging diagnosis. Respir Care 2008;53(10):1368–71.
[7] Houston J, Fleet M, Cowan M, McMillan N. Comparison of ultrasound with fluoroscopy in the assessment of suspected hemidiaphragmatic movement abnormality. Clin Radiol 1995;50(2):95–8.
[8] Lloyd T, Tang Y, Benson M, King S. Diaphragmatic paralysis: the use of M mode ultrasound for diagnosis in adults. Spinal Cord 2006:505–8. doi:10.1038/sj.sc.3101889.

[9] Nafisa S, Messer B, Downie B, Ehilawa P, Kinnear W, Algendy S, Sovani M. A retrospective cohort study of idiopathic diaphragmatic palsy: a diagnostic triad, natural history and prognosis. ERJ Open Res 2021;8(1). doi:10.1183/23120541.00953-2020.

[10] Alexander C. Diaphragm movements and the diagnosis of diaphragmatic paralysis. Clin Radiol 1966;17(1):79–83. doi:10.1016/S0009-9260(66)80128-9.

[11] Boussuges A, Gole Y, Blanc P. Diaphragmatic motion studied by m-mode ultrasonography: methods, reproducibility, and normal values. Chest 2009;135(2):391–400. doi:10.1378/chest.08-1541.

[12] Fayssoil A, Nguyen L, Ogna A, Stojkovic T, Meng P, Mompoint D, et al. Diaphragm sniff ultrasound: normal values, relationship with sniff nasal pressure and accuracy for predicting respiratory involvement in patients with neuromuscular disorders. Plos One 2019;14(4):e0214288. doi:10.1371/journal.pone.0214288.